\documentstyle[11pt,paspconf,epsf]{article}

\def\edcomment#1{\iffalse\marginpar{\raggedright\sl#1\/}\else\relax\fi}

\reversemarginpar

\begin{document}

\title{Extended CIII]$\lambda$1909 emission in Q0957+561}

\author{E. Mediavilla (IAC); V. Motta (IAC)\altaffilmark{1}; S. Arribas (IAC); E.E. Falco (CfA); A. Oscoz (IAC); M. Serra-Ricart (IAC); D. Alcalde (IAC); L. Goicoechea (UC); M. Ramella (OT); R. Barrena (IAC)}

\altaffiltext{1}{Departamento de Astronom\'{\i}a, Facultad de Ciencias, Universidad de la Rep\'ublica, Montevideo, Uruguay}

\begin{abstract}

2D spectroscopy of Q0957+561 in the CIII]$\lambda$1909 emission line reveals a rich variety of spatially extended features. We point out: (i) a blueshifted region apparently connecting the A and B compact images; (ii) an almost complete arc in the West side which fits well with the infrared arc detected by the Hubble Space Telescope and (iii) traces of extended emission in the East side which could be the fragments of a larger arc-shaped structure. These features should be explained by the current lens models.
\end{abstract}

Very recently, extended images of the host galaxies of QSOs have been detected in multiple imaged systems. Arcs of emission related to the galaxy hosts of several QSOs appear in infrared images taken with the Hubble Space Telescope (Impey et al 1998, CASTLEs). 2D spectroscopy of Q2237+0305 (Mediavilla et al. 1998) also shows an incomplete ring of emission associated to the CIII]$\lambda$1909 line.

\begin{figure}
\plotfiddle{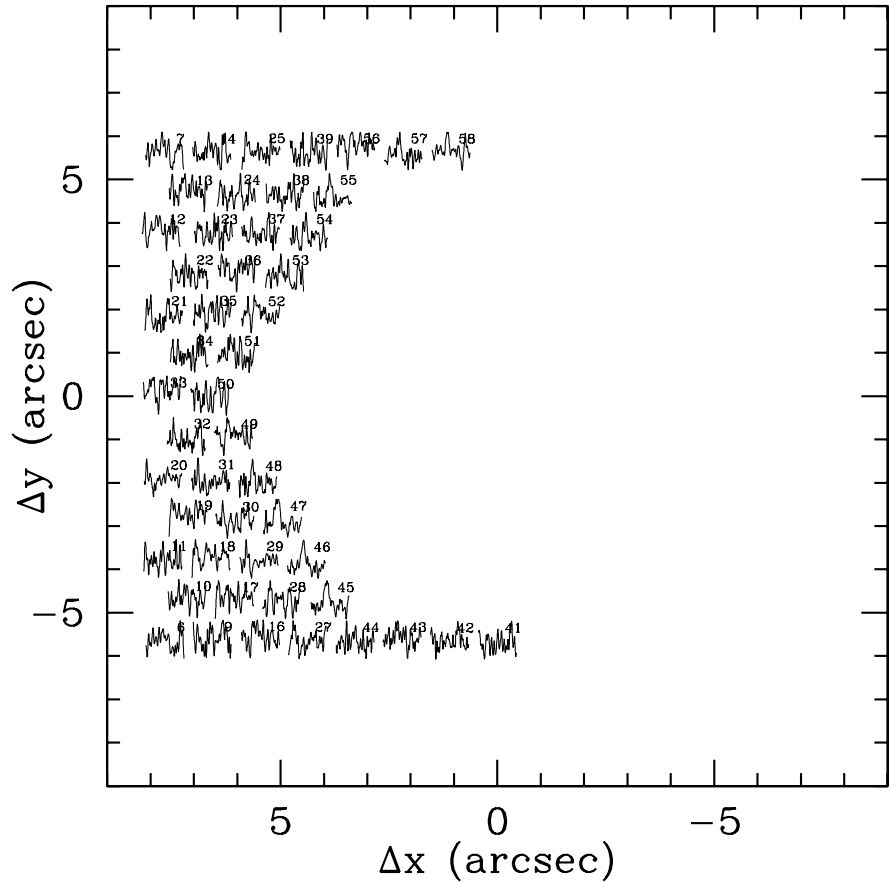}{9.2cm}{0}{60}{60}{-210}{-90}
\caption{2D distribution of spectra in $\lambda \, \sim \, 4300 \, - \, 5000$.}
\end{figure}

The spectra were taken the night of the 26 of December 1998, in INTEGRAL guaranteed time. We obtained 4 frames of 1800 seconds each with the SB2 fiber bundle of INTEGRAL (Arribas et al. 1998). After a standard reduction procedure we obtain a spectra for each fiber location on the sky (figure 1).

 To obtain a CIII]$\lambda$1909 line intensity map we have integrated the spectra after continuum substraction in the $\sim$ 4500-4700 wavelength range. In the resulting map (not shown here) we can see the two compact components but also a great deal of extended emission. We can point out several features: (i) the bridge apparently connecting the A and B images, (ii) an almost complete arc in the West side of the system, and (iii) possible fragments of another arc shaped structure towards the East. The West arc fits well with the incomplete ring detected in the HST infrared images of Q0957+561 (CASTLEs). However, the arc shaped feature tentatively detected in the East side would have dimensions greater than expected and would not be a symmetrical counterpart of the West arc.

To derive the {\em redshift} of the emitters we have performed Gaussian fittings to the CIII]$\lambda$1909 emission line. In spite of the manifest complexity of the line profile (possible causes are strong absorption or the presence of more than one system), we have fitted a unique Gaussian function in this preliminary analysis.  Inspection of the resulting map reveals a clear dependence of {\em redshift} with spatial location. It is also noticeable that the compact image centers are not at the redshift of the QSO (z=1.41). However, the line profiles of the spectra corresponding to these locations are not symmetrical and the Gaussian centroid could not represent the true redshift of a unique source. In any case, it is worth to mention that the line profile peaks correspond to z=1.41. The bridge apparently connecting A and B appears strongly blueshifted.

\end{document}